\documentclass[prl,twocolumn,superscriptaddress,longbibliography]{revtex4-2}
\usepackage{graphicx,amsmath,amssymb,tikz,scalerel,mathrsfs,hyperref,comment}
\usepackage{soul}
\usepackage{color, xcolor}
\hypersetup{
        colorlinks=true,       
        linkcolor=blue,        
        citecolor=blue,     
        urlcolor=blue}
\renewcommand{\section}[1]{{\par\it #1---}\ignorespaces}
\usetikzlibrary{svg.path}
\definecolor{orcidlogocol}{HTML}{A6CE39}
\tikzset{
	orcidlogo/.pic={
		\fill[orcidlogocol] svg{M256,128c0,70.7-57.3,128-128,128C57.3,256,0,198.7,0,128C0,57.3,57.3,0,128,0C198.7,0,256,57.3,256,128z};
		\fill[white] svg{M86.3,186.2H70.9V79.1h15.4v48.4V186.2z}
		svg{M108.9,79.1h41.6c39.6,0,57,28.3,57,53.6c0,27.5-21.5,53.6-56.8,53.6h-41.8V79.1z M124.3,172.4h24.5c34.9,0,42.9-26.5,42.9-39.7c0-21.5-13.7-39.7-43.7-39.7h-23.7V172.4z}
		svg{M88.7,56.8c0,5.5-4.5,10.1-10.1,10.1c-5.6,0-10.1-4.6-10.1-10.1c0-5.6,4.5-10.1,10.1-10.1C84.2,46.7,88.7,51.3,88.7,56.8z};}}
\newcommand\orcid[1]{\href{https://orcid.org/#1}{\mbox{\scalerel*{\begin{tikzpicture}[yscale=-1,transform shape]\pic{orcidlogo};\end{tikzpicture}}{|}}}}
\UseRawInputEncoding
\usepackage[]{lineno}

\let\oldequation\equation
\let\oldendequation\endequation

\renewenvironment{equation}
{\linenomathNonumbers\oldequation}
{\oldendequation\endlinenomath}

\begin{document}
\setlength\linenumbersep{0.15cm} %
\title{Impact of noise on nonlinear-exceptional-point-based sensors}
\author{Kai Bai}
\affiliation{Key Laboratory of Artificial Micro- and Nano-structures of Ministry of Education and School of Physics and Technology, Wuhan University, Wuhan 430072, China}
\author{Chen Lin}
\affiliation{Key Laboratory of Artificial Micro- and Nano-structures of Ministry of Education and School of Physics and Technology, Wuhan University, Wuhan 430072, China}
\author{Meng Xiao}
\email{phmxiao@whu.edu.cn}
\affiliation{Key Laboratory of Artificial Micro- and Nano-structures of Ministry of Education and School of Physics and Technology, Wuhan University, Wuhan 430072, China}
\affiliation{Wuhan Institute of Quantum Technology, Wuhan 430206, China}

\begin{abstract}
Nonlinear exceptional points (NEPs), a new type of spectral singularity in nonlinear non-Hermitian systems, are expected to address the noise divergence issue encountered at linear exceptional points and are therefore under the scrutiny of theoretical and experimental investigations. However, concerns have been raised that NEPs may hinder improvements in the signal-to-noise ratio (SNR) of sensors, and there is currently no rigorous theoretical framework to characterize noise effects in NEPs, particularly when accounting for the inherent nonlinear feedback. Here, we develop a new theoretical framework to address the impact of noise on NEP-based sensors, effectively resolving these concerns. The interplay between noise and nonlinearity keeps the average frequency virtually unchanged. In addition, a hidden feedback mechanism limits the increase in detectable uncertainty, together enabling a substantial SNR enhancement at NEPs. Our results resolve the ongoing debate over the SNR of NEPs and lay the groundwork for NEP-based sensor technologies.
\end{abstract}

\maketitle
\section{Introduction}\label{introduction}
Sensing plays a significant role in driving innovation in science and technology, with applications ranging from the analysis of biomolecules and nanoparticles \cite{1science1145002,2pnas0808988106,3Zhu2010a}, the measurement of superconducting qubits \cite{4RevModPhys.82.1155,5PhysRevA.69.062320}, to wireless telemetry \cite{6Chen2018b,7Dong2019NE}, health monitoring \cite{8Chen2014a}, and gravitational wave detection \cite{9RBallantini_2003}. Most sensors rely on ubiquitous dispersive measurements, where a parameter of interest shifts the frequency of a resonance or splits the spectrum of two degenerate modes \cite{1science1145002,2pnas0808988106,3Zhu2010a,4RevModPhys.82.1155,5PhysRevA.69.062320,9RBallantini_2003,10N5299019,11C967381,12C4660282}. With recent advances in non-Hermitian physics \cite{13Ashida02072020,14Heiss_2012,15Rotter_2009,16PhysRevLett.133.180801}, there has been considerable current interest in sensors based on exceptional points (EPs) \cite{17PhysRevA.96.033842,18Hodaei2017,19PhysRevLett.128.173602,20PhysRevX.6.021007,21Wang2019,22Tang,23PhysRevB.99.241403,24PhysRevLett.132.243601,25PhysRevLett.132.243802,26PhysRevLett.133.133601,27PhysRevLett.112.203901,28PhysRevLett.117.110802,29PhysRevA.93.033809,30Chen2017,31PhysRevLett.125.203602,32PhysRevLett.123.213901,33PhysRevLett.128.203904,34PhysRevLett.127.186601}. A small perturbation to the system Hamiltonian at an EP leads to a significant shift in the resonance frequency. This enhanced frequency shift, when divided by the small perturbation, defines the responsivity and presents a promising approach to improving sensing precision. However, due to the loss of completeness of the eigenbasis \cite{35PhysRevA.98.023805,36Chen_2019,37Lau2018,38Wang2020,39PhysRevLett.123.180501}, the unavoidable noise is also dramatically increased near the EP. As a result, the corresponding signal-to-noise ratio (SNR), which scales linearly with the responsivity over the minimal measurable uncertainty, does not improve in the vicinity of EPs \cite{35PhysRevA.98.023805,36Chen_2019,37Lau2018,38Wang2020,39PhysRevLett.123.180501}. 

To improve the corresponding SNR \cite{40PhysRevLett.130.227201,41Kononchuk2022}, nonlinear exceptional points (NEPs) \cite{42Bai2023,43PhysRevLett.132.073802,44PhysRevLett.130.266901}, novel spectral singularities in nonlinear non-Hermitian systems, have been proposed to address the noise issues encountered at linear EPs. At NEPs, certain steady self-consistent eigenmodes (which may include auxiliary modes) of the nonlinear Hamiltonian coalesce, i.e., collapse into a single state. This collapse of eigenmodes ensures that the key spectral features of conventional linear EPs, such as enhanced responsivity, are retained. Intriguingly, the instantaneous Hamiltonian governing the system's dynamics can possess a complete eigenbasis, as evidenced by a finite Petermann factor (PF), in contrast to the divergent PF at linear EPs \cite{38Wang2020,45PhysRevA.78.015805,46PhysRevA.39.1253}. The complete eigenbasis and the feedback mechanism inherent in nonlinear systems are expected to offer a promising pathway to mitigate the noise-related challenges. In addition, compared to a conventional linear order-$n$ EP ($\text{EP}_n$, where $n \in \mathbb{N}$), an order-$n$ NEP ($\text{NEP}_n$) is much easier to achieve, as the number of required constraints is significantly reduced  \cite{44PhysRevLett.130.266901}. Consequently, NEPs are highly desirable for many EP-related applications, such as signal detection \cite{18Hodaei2017,30Chen2017}, chiral state transfer \cite{47Xu2016,48PhysRevLett.118.093002,CSTBai2025}, and EP-based on-chip optical devices \cite{49Zhang2019,50Choi2017,51Peng2016}, and are particularly significant in sensing. However, recent concerns have emerged that the interplay between noise and nonlinearity may introduce previously unexplored obstacles, potentially hindering enhanced sensing capabilities at NEPs \cite{52PhysRevLett.134.133801}. Specifically, the unavoidable noise may shift the positions of NEPs in the parameter space and reduce their order, thereby eliminating the anticipated divergence in the SNR of NEP-based sensors. Clearly, resolving this debate is critically important, not only for advancing the theoretical understanding of the interplay between nonlinearity and noise near NEPs but also for driving innovation in NEP-based sensing technologies.

Here, we address the two core questions outlined above: (1) What is the underlying mechanism governing the interplay between nonlinearity and noise? (2) Can operating near a NEP enhance the SNR of a sensor? To answer these questions, we analyze an exemplary scheme of $\text{NEP}_3$ sensing based on two coupled resonators with a nonlinear saturable gain. Near the $\text{NEP}_3$, the response of eigenfrequencies demonstrates a third-order root law, so that the system can potentially act as a NEP sensor. In a noisy environment, we develop a framework that exactly captures the fluctuations around the steady state. Specifically, these fluctuations are governed by an effective nonlinear Langevin equation with an embedded hidden feedback mechanism, which confines their amplitude to remain finite and consequently results in a finite measurement uncertainty. We extract the average frequency and the associated uncertainty from the phase accumulation during the dynamical evolution. Near the $\text{NEP}_3$, the average frequency still follows a third-order root law and the uncertainty remains bounded. As a result, a significant enhancement in SNR is achieved. Our results advance the theoretical understanding of noise in nonlinear systems and take essential steps toward resolving the ongoing debate on the SNR of NEPs, thereby laying the foundation for the development of NEP-based sensor technologies.

\begin{figure}
	\centering
	\includegraphics[width=1\columnwidth]{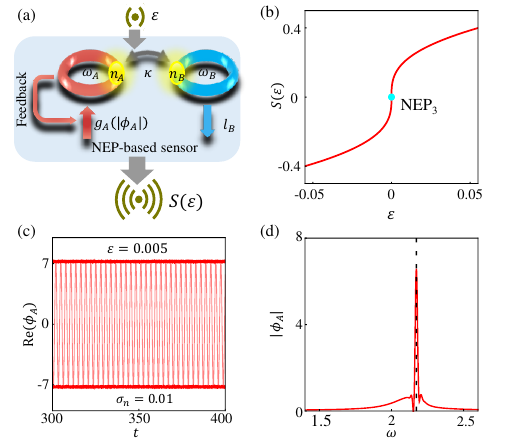}\\
	\caption {(a) Schematic of the workflow of the NEP-based sensor. Within the sensor’s range, a perturbation induces a small change $\varepsilon$ in the sensor, which further gives rise to a significant variation in a detectable quantity $S(\varepsilon)$. $n_{A,B}$ represent the noise term. (b) Retrieved $S(\varepsilon)$ as a function of $\varepsilon$ when the noise terms are ignored. (c) A segment of the evolution of $\text{Re}(\phi_A)$ in a noisy environment, showing the situation after the system reaches a stable state. The initial state is set to $\phi_A (0)=7$ and $\phi_B (0)=-7i$, and $\varepsilon=0.005$. (d)  Average spectrum obtained via fast Fourier transform over an evolution time of $T_0=2^{22}\times 10^{-4}$, averaged over 1000 independent simulations. The black dashed line indicates the corresponding steady-state eigenfrequency $\omega_s$ at $\varepsilon=0.005$. Other parameters used in (b-d) are: $\kappa=l_B=1$, $g_0=\omega_B=\omega_A=2$, $\Gamma=0.02$, and $\sigma_n = 0.01$. }
	\label{fig1}
\end{figure}
\section{NEP sensing}
We consider a generic noisy NEP-based sensor perturbed by a small parameter $\varepsilon$. The schematic of the workflow is shown in Fig.~\ref{fig1}(a). The system dynamics are governed by the time-dependent nonlinear Schr\"odinger equation: 
\begin{equation}
id\left|\boldsymbol{\phi}\right\rangle /dt=H_{\left|\boldsymbol{\phi}\right\rangle }\left|\boldsymbol{\phi}\right\rangle, \label{eq1}
\end{equation}
where $\left|\boldsymbol{\phi}\right\rangle\equiv(\phi_A,\phi_B)^T$ is the state vector, with the superscript  $T$ denoting transpose, and $\phi_A$ and $\phi_B$  represent the complex mode amplitudes of the red (left) and blue (right) resonators, respectively. The nonlinear Hamiltonian $H_{\left|\boldsymbol{\phi}\right\rangle}$  that depends on $\left|\boldsymbol{\phi}\right\rangle$  is given by
\begin{equation}
H_{\left|\boldsymbol{\phi}\right\rangle }=\left(\begin{array}{cc}
\omega_{A}+\varepsilon+ig_{A}(|\phi_{A}|) & \kappa\\
\kappa & \omega_{B}-il_{B}
\end{array}\right), \label{eq2}
\end{equation}
where $\omega_A (\omega_B)$ is the resonant frequency, $\kappa$ is the coupling strength, $l_B$ denotes a linear loss, and $g_A (|\phi_A |)$ denotes a nonlinear saturable gain that decreases with increasing field amplitude $|\phi_A |$. Without loss of generality, the gain saturation in this work is modeled by the van der Pol form $g_A (|\phi_A |)=g_0+\Gamma(1-|\phi_A |^2)$ \cite{52PhysRevLett.134.133801,54PhysRevLett.123.250401,55PhysRevLett.111.234101}. And for simplicity, all the parameters are normalized by $\kappa$. A $\text{NEP}_3$ can be achieved at $\kappa=l_B=1$ and $\omega_B=\omega_A$ [see Supplemental Material, Sec. 1]. The unperturbed Hamiltonian is configured at this $\text{NEP}_3$, and we assume that the parameter to probe introduces a slight shift $\varepsilon$ in the resonant frequency of the left resonator, i.e., $\omega_A$. 

The measurable quantities include eigenfrequency variation, as well as the relative amplitude change and phase difference change of eigenstates [see Fig. S1]. Given the widespread use of spectral measurement schemes, we take the eigenfrequency variation, denoted as $S(\varepsilon)\equiv\omega-\omega_{0}$, as the measurable quantity of the sensor. Here, $\omega_0$ denotes the frequency of $\text{NEP}_3$, which equals $2$ for the parameters chosen above, while $\omega$ represents the eigenfrequency of the sensor under the perturbation $\varepsilon$. $\omega$ can be obtained either by performing a Fourier transform of the state evolution or by recording the accumulated phase during the dynamical evolution [see Fig. S2]. Figure \ref{fig1}(b) shows $S(\varepsilon)$ as a function of $\varepsilon$ without noise, exhibiting an $\varepsilon^{1/3}$ scaling behavior as $\varepsilon$ approaches 0. Consequently, the sensor's responsivity to the perturbative strength, $\partial S/\partial\varepsilon \propto \varepsilon^{-2/3}$ --- which diverges as $\varepsilon \to 0$ --- is significantly enhanced compared to that of non-EP and linear EP systems composed of two resonant modes. 

\begin{figure}
	\centering
	\includegraphics[width=1\columnwidth]{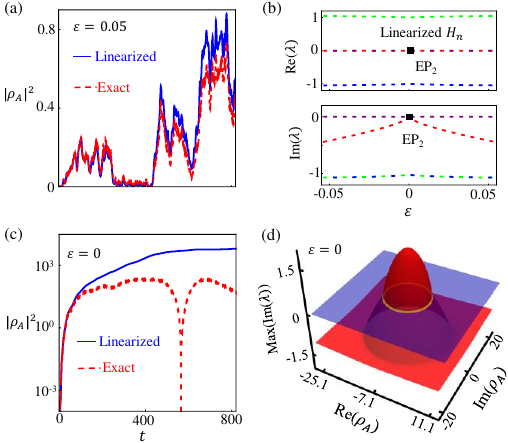}\\
	\caption {(a) Evolution of $|\rho_A|^2$ for $\varepsilon=0.05$. (b) The eigenfrequencies of $\tilde{H}_{eff}$ versus $\varepsilon$. At $\varepsilon=0$, an $\text{EP}_2$ emerges, as marked by the black square. (c) Evolution of $|\rho_A|^2$ for $\varepsilon=0$. (d) The maximum value of the imaginary parts of all four eigenvalues of $H_n$ as a function of $\rho_{A}$. The blue plane corresponds to a value of zero. In (a, c), the blue lines and red dashed lines correspond to the linearized approximation and the exact simulation, respectively. The initial states are set to $\rho_{A,B}=0$, and the details of the white noise sequences are provided in the Supplemental Material, Sec. 4. In (d), $\omega_r=\omega_s$ and $A_r=A_s$, and all the other parameters used are the same as those in Fig.~\ref{fig1}.}
	\label{fig2}
\end{figure}
\section{Noise analysis}\label{Noise-analysis}
While the enhanced responsivity at the $\text{NEP}_3$ provides a promising avenue for advancing sensing performance, several challenges persist. On the one hand, unlike in linear non-Hermitian systems where noise enhancement can be characterized by the Petermann factor, the situation in nonlinear systems is far more intricate due to the nonlinear interplay between system and noise [see Supplemental Material, Secs. 2 and 3]. At present, the anticipated superior performance of NEP sensors in noisy environments lacks rigorous theoretical validation. On the other hand, recent concerns suggest that the interplay of noise and nonlinearity introduces previously unidentified obstacles to enhanced sensing \cite{52PhysRevLett.134.133801}. To address this ongoing debate, a quantitative characterization of the impact of nonlinearity and noise interaction near a NEP on sensor performance is essential. To this end, we develop a theory to evaluate the performance of a NEP sensor. The (quantum) Langevin equation for the NEP sensor is given by
\begin{equation}
id\left|\boldsymbol{\phi}\right\rangle /dt=H_{\left|\boldsymbol{\phi}\right\rangle }\left|\boldsymbol{\phi}\right\rangle+(n_A,n_B)^T. \label{eq3}
\end{equation}
Here, $n_x$ with $x=A,B$ represent white noise that has zero mean and satisfy $\langle n_{A}^{*}(t)n_{B}(t^{\prime})\rangle=\sigma^{2}_n\delta_{A,B}\delta(t-t^{\prime})$ and $\langle n_{A}(t)n_{B}(t^{\prime})\rangle=0$. The noise can originate from vacuum fluctuations, thermal noise associated with the loss and gain components. or other possible sources. The validity of Eq.~\eqref{eq3} has been confirmed through quantitative agreement with experiments and full quantum calculations \cite{56PhysRevLett.129.013901,57RevModPhys.85.299}. 

The simulations are implemented in Mathematica, and the noise is modeled as a Wiener process \cite{58Andreas}. Figure S4 calibrates the statistical discrete probability distribution, the correlation function, and the spectrum of the randomly generated white noise. For each independent simulation, we solve Eq. \eqref{eq3} with a fixed-step stochastic Runge-Kutta method starting from a fixed initial state. After a short time, field amplitudes become nearly constant, indicating that the system has reached a stable state [see Fig. S5]. Figure \ref{fig1}(c) shows a segment of a typical stochastic evolution of $\text{Re}(\phi_A)$ after the system reaches a stable state for $\varepsilon=0.005$, where the solid red lines are the envelope of $\text{Re}(\phi_A)$. By performing a Fourier transform over the entire simulation time $T_0=2^{22}\times10^{-4}$, we obtain the average spectrum from 1000 independent simulations, shown in Fig.~\ref{fig1}(d). Here, only one spectral peak is observed, which matches well with the black dashed line denoting the steady-state eigenfrequency $\omega_s$ obtained in the absence of noise. A zoomed-in view [Fig. S6] clearly shows that the peak frequencies, $\omega_c$ and $\omega_s$, align perfectly within the resolution of the FFT. We note that, unlike Ref.~\cite{52PhysRevLett.134.133801}, here we focus on the regime where the noise level is small. As the noise level increases, the FFT spectrum begins to exhibit double peaks, and this onset of peak splitting occurs at progressively lower noise levels when approaching the $\text{NEP}_3$ [Fig. S12]. 

To rigorously characterize the interaction between nonlinearity and noise and to uncover the underlying mechanisms, we expand $\phi_A(t)$ and $\phi_B(t)$ as follows: $\phi_A \equiv \psi_A e^{-i\omega_r t} \equiv (A_r + \rho_A)e^{-i\omega_r t}$ and $\phi_B \equiv \psi_B e^{-i\omega_r t} \equiv (B_r + \rho_B)e^{-i\omega_r t}$, respectively. Here, $\omega_r$ is an arbitrary real number, not restricted to the eigenfrequency $\omega_s$ or the peak frequency $\omega_c$, $A_r$ is a complex constant, and $B_r \equiv A_r B_s / A_s$, with $(A_s, B_s)^T$ denoting the stable state without noise. The quantities $\rho_{A,B}$ represent time-dependent oscillatory components. Notably, here the amplitudes of $\rho_{A}$ and $\rho_{B}$ are not required to be smaller than those of the complex constants $A_r$ and $B_r$. Consequently, the nonlinear gain $g_A(|\psi_A|)$ becomes $g_A(|A_r|^2) + \delta_g$, where $\delta_g \equiv -\Gamma(|\rho_A|^2 + \rho_A^* A_r + \rho_A A_r^*)$. To obtain the dynamics of $\rho_{A,B}$, we further employ the relation 
\begin{eqnarray}
&&\left(\begin{array}{cc}
\bar{\omega}_{A}+ig_{s}-\omega_{s}+\omega_{r} & \kappa\\
\kappa & \bar{\omega}_{B}-il_{B}-\omega_{s}+\omega_{r}
\end{array}\right)\left(\begin{array}{c}
A_{r}\\
B_{r}
\end{array}\right)\nonumber\\&&=0, \label{eq4}
\end{eqnarray}
 where $\bar{\omega}_A \equiv \omega_A + \varepsilon - \omega_r$, $\bar{\omega}_B \equiv \omega_B - \omega_r$, and $g_s$ denotes the saturated value of the gain in the absence of noise. A direct calculation yields (see Supplemental Material Sec. 4 for details) \begin{eqnarray}
\dot{\rho}_{A}&=&X\rho_{A}+Y\rho_{A}^{*}+C\rho_{B}+E-i\zeta_{A}\nonumber,\\ \dot{\rho}_{B}&=&C\rho_{A}+D\rho_{B}+K-i\zeta_{B}, \label{eq5}
\end{eqnarray}
where $X=-i\bar{\omega}_{A}+g_{A}(|A_{r}+\rho_{A}|^{2})-\Gamma|A_{r}|^{2}$, $Y=-\Gamma|A_{r}|^{2}$, $C=-i\kappa$, $D=-i\bar{\omega}_{B}-l_B$, $E=(g_{A}(|A_{r}|^{2})-g_{s}-\Gamma|\rho_{A}|^{2}-i(\omega_{s}-\omega_{r}))A_{r}$, $K=-i(\omega_s-\omega_r) A_r B_s/A_s$, and $\zeta_{A,B}=n_{A,B} e^{i\omega_{r}t}$. To streamline the representation, we define the vectors $|\rho\rangle\equiv(\rho_A,\rho_A^*,\rho_B,\rho_B^* )^T$, $|\zeta\rangle\equiv(-i\zeta_A,i\zeta_A^*,-i\zeta_B,i\zeta_B^* )^T$, and $|M\rangle\equiv(E,E^*,K,K^* )^T$.  Accordingly, the dynamics of $|\rho\rangle$ can be written as
\begin{eqnarray}
i|\dot{\rho}\rangle=H_n |\rho\rangle+i|M\rangle+i|\zeta\rangle, \label{eq6}
\end{eqnarray}
where the Hamiltonian $H_n$ has the form
\begin{eqnarray}
H_{n}=i\left(\begin{array}{cccc}
X & Y & C & 0\\
Y^{*} & X^{*} & 0 & C^{*}\\
C & 0 & D & 0\\
0 & C^{*} & 0 & D^{*}
\end{array}\right). \label{eq7}
\end{eqnarray}
Note that $H_n$ is inherently nonlinear because the term $X$ contains contributions from $\rho_A$. We need to emphasize that no approximation was made in the derivation of Eq.~\eqref{eq6}. When $|\varepsilon|\gg0$, it turns out that the dynamics of $|\rho\rangle$
can be well captured by a linearized $H_n$ (defined as $\tilde{H}_{eff}$) under the first-order approximation, where we set $g_A = g_s$, $\omega_r = \omega_s$, and consequently $E = K = 0$. As a numerical demonstration, Fig.~\ref{fig2}(a) compares the evolution of $|\rho_A|^2$ governed by the linearized equations (the blue line, replacing $H_n$ with $\tilde{H}_{eff}$ and setting $|M\rangle=0$), with the exact numerical simulations (the red dashed line) at $|\varepsilon|=0.05$. The close agreement between these two lines over a long enough time demonstrates the validity of the linearized approximation. Since $\tilde{H}_{\text{eff}}$ is already linearized and can accurately predict the system dynamics, as demonstrated in Fig.~\ref{fig2}(a), the noise amplification can be treated as proportional to the PF of the eigenstates of $\tilde{H}_{\text{eff}}$ \cite{52PhysRevLett.134.133801}. Figure~\ref{fig2}(b) shows the four eigenvalues of $\tilde{H}_{eff}$. Notably, at $\varepsilon=0$, an $\text{EP}_2$ emerges, and the corresponding PF diverges [see Fig. S8(c)]. As a result, one would naturally expect $|\rho\rangle$ to diverge over long times due to the presence of noise; however, this is not the case.

In fact, as $\varepsilon \to 0$, the dynamics of $|\rho\rangle$ under exact numerical simulations, i.e., Eq.~\eqref{eq6} without linearization, increasingly deviate from the predictions based on $\tilde{H}_{\text{eff}}$. Figure~\ref{fig2}(c) displays the exact evolution trajectory of $|\rho_A|^2$ (the red dashed line) at $\varepsilon=0$, which remains bounded and deviates from the linearized approximation (the blue line). Here, the prediction that the noise amplification can be characterized by the PF of $\tilde{H}_{eff}$ fails due to the feedback mechanism imposed by the nonlinear saturable gain. To further elucidate this hidden feedback mechanism that confines the noise effects, the red surface in Fig.~\ref{fig2}(d) represents the maximum imaginary part among all eigenvalues of $H_n$, denoted $\max(\text{Im}(\lambda))$, plotted against $\rho_A$. For reference, the transparent blue surface represents zero, intersecting the red surface at the yellow circle. Inside the yellow circle, $\max(\text{Im}(\lambda))$ is positive, indicating exponential growth in $\rho_A$. When $|\rho_A|$ grows sufficiently large to extend beyond the yellow circle in Fig.~\ref{fig2}(d), $\max(\text{Im}(\lambda))$ becomes negative, indicating that $|\rho_A|$ subsequently undergoes exponential decay. Consequently, as long as the noise term $|\zeta\rangle$ remains finite, this feedback mechanism constrains $|\rho_A|$ to be bounded. To proceed, we expand $\rho_A \equiv \sum_\omega \alpha_\omega e^{-i\omega t}$ and evaluate the long-time average, yielding $\langle|\rho_A|^2\rangle = \sum_\omega |\alpha_\omega|^2$, where the time average is defined as $\langle\cdot\rangle = \int_0^{T_s} (\cdot)\, dt / T_s$ with $T_s \to \infty$.  As $|\rho_A|$ remains finite, the effect of noise, which is contained in $|\alpha_\omega|$, is also finite, thereby ensuring a finite linewidth.

\begin{figure}
	\centering
	\includegraphics[width=1\columnwidth]{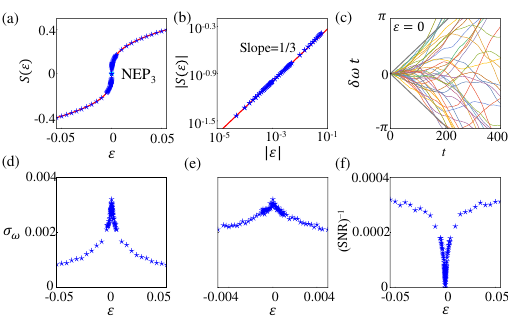}\\
	\caption {(a) The blue stars represent the ensemble-averaged $S(\varepsilon)$ as a function of $\varepsilon$, which coincides with the solid red line obtained in the absence of noise. (b) Double-logarithmic plot of $|S(\varepsilon)|$ versus $|\varepsilon|$, where the solid line denotes a reference line with a slope of $1/3$. (c) Time evolution of the accumulated phase $\delta_{\omega} t$  for 40 independent noise realizations. (d) The extracted frequency uncertainty $\sigma_{\omega}$ as a function of $\varepsilon$. (e) Near $\varepsilon \approx 0$, the linearized approximation $\tilde{H}_{eff}$ breaks down, and $\sigma_{\omega}$ converges to a finite value. (f) SNR as a function of $\varepsilon$, showing the expected significant enhancement near the $\text{NEP}_3$. All results are obtained by solving Eq.~\eqref{eq3}. The initial conditions are set to $\omega_r = \omega_s$, $A_r = A_s$, and $\rho_{A,B} = 0$,  with all other parameters identical to those used in Fig.~\ref{fig1}. The ensemble average is computed over 1000 realizations for the results in (a), (b), and (d-f).}
	\label{fig3}
\end{figure}

To demonstrate the SNR enhancement in NEP-based sensing schemes, we extract $S(\varepsilon)$ from the phase accumulation during dynamical evolution, which can be directly measured using an oscilloscope or equivalent techniques \cite{42Bai2023,43PhysRevLett.132.073802,59Chen2024}. We focus on the phase of $\phi_A(t)$, denoted as $\theta_A(t)$. A total of 1000 independent simulations are performed, each lasting for a duration of $\delta t = 13$, corresponding to roughly four oscillation periods near $\varepsilon = 0$. The frequency in each run is extracted as follows:
\begin{eqnarray}
S(\varepsilon) = -\frac{1}{\delta t} \int_0^{\delta t} d\theta_A(t) \equiv \omega_s + \delta_\omega, \label{eq8}
\end{eqnarray}
where $\delta_\omega$ denotes the frequency deviation, and its dependence on $\varepsilon$ is omitted for clarity.  Figure \ref{fig3}(a) plots $S(\varepsilon)$ (blue stars) as a function of $\varepsilon$, which coincides with the ideal values (solid red line).  Figure \ref{fig3}(b) shows $|S(\varepsilon)|$ versus $|\varepsilon|$ on a logarithmic scale, where the slope fits well to $1/3$ even for extremely small $|\varepsilon|$, revealing a significant enhancement in responsivity. Figure \ref{fig3}(c) presents the phase residuals $\delta_\omega t$ for 40 independent realizations at $\varepsilon=0$, which are almost confined within two gray lines corresponding to $\pm 2\pi t/T_0$.  In other words, in our $\text{NEP}_3$-based sensor, the noise-induced variation $\delta_\omega$ lies within the resolution limit of the FFT, and remains bounded even at the $\text{NEP}_3$. Figure~\ref{fig3}(d) shows the corresponding frequency uncertainty $\sigma_\omega$, defined as the standard deviation of $S(\varepsilon)$. When $\varepsilon$ is sufficiently far from the $\text{NEP}_3$, the linearized approximation $\tilde{H}_{eff}$ remains valid. As $\varepsilon$ approaches zero, the frequency uncertainty $\sigma_\omega$ gradually increases. However, as $\varepsilon$ is close enough to zero, the linearized approximation fails, and $\sigma_\omega$ converges to a finite value [see Fig.~\ref{fig3}(e)]. As a result, the SNR, proportional to $(\partial S(\varepsilon)/\partial \varepsilon)/\sigma_\omega$ (assuming a proportionality factor of 1), can still be dramatically improved near the $\text{NEP}_3$. Further discussion on the impact of noise on SNR enhancement is provided in the Supplementary Material, Sec. 5. Notably, at $\sigma_n=0.01$, the frequency uncertainty induced by noise $\sigma_\omega$ is smaller than the precision of the FFT and is typically less than the errors caused by experimental materials, environmental factors, and other uncertainties. Therefore, in practical experiments, $\sigma_\omega$ is predominantly limited by the intrinsic resolution of the measurement instruments.
\section{Summary and Discussions}\label{conclusion}
In summary, we have established a theoretical framework based on nonlinear Langevin equations to investigate the interplay between noise and nonlinearity, and have demonstrated the enhancement in SNR of NEP-based sensors. The ensemble-averaged $S(\varepsilon)$ aligns well with the eigenfrequency of the noise-free nonlinear non-Hermitian system, and the standard deviation remains very small within the range of interest due to the hidden feedback mechanism. These results contribute to a substantial enhancement in the SNR. Thus, our work resolves the SNR debate in NEP-based sensors, enables applications in noisy environments, and can potentially extend the NEP framework to condensed matter and open quantum systems.

Unlike previous works \cite{52PhysRevLett.134.133801,20PhysRevX.6.021007,22Tang} relying on frequency peaks as measurable quantities, we instead detect the phase accumulation over a finite time interval. This approach withstands significantly stronger noise levels, remaining effective even when the single-frequency peak is washed out (see Supplemental Material, Sec. 5). It is important to note that, similar to EP-based sensors, NEP sensors also require the signal to persist for a sufficient duration to allow the system to reach a stable state. Exploring ways to design and harness noise to reduce this stabilization time presents an exciting direction for future research. In addition, the PF of the instantaneous Hamiltonian that was employed in Refs.~\cite{42Bai2023,43PhysRevLett.132.073802,44PhysRevLett.130.266901} only provides an exaggerated (and unattainable) noise enhancement factor, leading to an underestimated SNR enhancement [see Supplemental Material, Sec. 3]. 

\begin{acknowledgments}
The authors would like to thank Dr. X. Zheng and Prof. Y. D. Chong for helpful discussions. This work is supported by the National Key Research and Development Program of China [Grant No. 2022YFA1404900], the National Natural Science Foundation of China [Grants No. 12274332, 12334015, 12321161645, 12404440], the China Postdoctoral Science Foundation under Grant No. 2023M742715, the China National Postdoctoral Program for Innovative Talents under Grant No. BX20240266, and Postdoctor Project of Hubei Province under Grant No. 2024HBBHCXB054.   
\end{acknowledgments}
\bibliography{reference}

\end{document}